\documentstyle[preprint,eqsecnum,aps]{revtex}
\newif\iftightenlines\tightenlinesfalse
\tightenlines\tightenlinestrue
\begin{document}
%
\def\pT{p_T^{\phantom{7}}}
\def\MW{M_W^{\phantom{7}}}
\def\ET{E_T^{\phantom{7}}}
\def\bh{\bar h}
\def\lm{\,{\rm lm}}
\def\lo{\lambda_1}
\def\lt{\lambda_2}
\def\pslt{p\llap/_T}
\def\eslt{E\llap/_T}
\def\to{\rightarrow}
\def\Re{{\cal R \mskip-4mu \lower.1ex \hbox{\it e}}\,}
\def\Im{{\cal I \mskip-5mu \lower.1ex \hbox{\it m}}\,}
\def\SU{SU(2)$\times$U(1)$_Y$}
\def\te{\tilde e}
\def\tl{\tilde\ell}
\def\tt{\tilde t}
\def\tb{\tilde b}
\def\ttau{\tilde \tau}
\def\tg{\tilde g}
\def\tga{\tilde \gamma}
\def\tnu{\tilde\nu}
\def\tell{\tilde\ell}
\def\tq{\tilde q}
\def\tw{\widetilde W}
\def\tz{\widetilde Z}
%
%
\preprint{\vbox{\baselineskip=14pt%
   \rightline{FSU-HEP-940204}\break
   \rightline{UH-511-781-94}
}}
\title{DETECTING HIGGS BOSON DECAYS TO NEUTRALINOS \\
AT HADRON SUPERCOLLIDERS}
\author{Howard Baer$^1$, Mike Bisset$^2$, Chung Kao$^1$ and Xerxes Tata$^2$}
\address{
$^1$Department of Physics,
Florida State University,
Tallahassee, FL 32306 USA
}
\address{
$^2$Department of Physics and Astronomy,
University of Hawaii,
Honolulu, HI 96822 USA
}
\date{\today}
\maketitle
\begin{abstract}
We examine prospects for detecting the neutral Higgs bosons of minimal
supersymmetric models (MSSM) when their decays into neutralino pairs are
kinematically allowed.
The best signature appears to be $H_h,H_p\to\tz_2\tz_2\to 4\ell +\eslt$.
We argue that Standard Model contributions
to this signature are negligible, and examine regions of MSSM parameter space
where
the four lepton mode should be observable at the Large Hadron Collider.
The same signal can also come from
continuum neutralino pair production. We propose a set of cuts to
illustrate that the neutralino decay mode of the Higgs bosons provides
a viable signal over a substantial range of
model parameters, and show that it may be separable from continuum
neutralino production if sufficient integrated luminosity can be accumulated.

\end{abstract}
\medskip
%
%
%
\section{Introduction}

The derivation of large radiative corrections to Higgs boson masses\cite{HMASS}
and couplings in the Minimal Supersymmetric Standard Model (MSSM)\cite{MSSM}
has led a number of groups\cite{BBKT,GUNION,BARGER,KZ,BBDKT} to
re-evaluate prospects for SUSY Higgs detection at various colliding
beam facilities. Most of these studies have focused on regions of
MSSM parameter space where various standard model (SM) decay modes
({\it e.g.} $\gamma\gamma$ and $ZZ$ or $ZZ^*\to 4\ell$ ) of
the SUSY Higgs bosons are detectable above background; parameter space choices
were selected such that SUSY particle masses were large so that
Higgs boson decays to sparticles were kinematically forbidden. A region of
parameter space roughly spanning pseudoscalar Higgs mass $m_{H_p}\sim
100-300$~GeV and ratio of Higgs vevs $\tan\beta\sim 4-10$ (for $\tan\beta >
10$, the observability of the signal is sensitive to the detectability of the
$\tau\bar{\tau}$ decay modes of the Higgs bosons\cite{UNAL}) was
found where {\it none} of the SM decay modes were detectable\cite{FN0}.
Very recently,
it has been argued\cite{DGV} that  Higgs bosons, produced in association
with $t$-quark pairs and identified via their dominant $b\bar{b}$ decays
may fill this ``hole", provided
that sufficient $b$-tagging capability can be achieved.

Over the last year or two, several groups\cite{SPECTRA} have studied
grand unified models within the supergravity
framework and have shown that it is quite possible to construct models
consistent
with constraints from colliders, proton decay experiments and
cosmology. Interestingly, these analyses generally find that the sparticles
are all considerably lighter than 1 TeV, and further, that the lighter chargino
($\tw_1$) and the two lighter neutralinos ($\tz_1$ and $\tz_2$) frequently have
masses in the range 50-150 GeV so that these may well be accessible via
the decays of the MSSM Higgs bosons.

In Ref. \cite{BBDKT}, we studied how these supersymmetric decay modes
would affect the
phenomenology of the Higgs sector. We showed that the chargino and
neutralino decays of MSSM Higgs bosons have substantial branching
fractions when kinematically allowed, and can sometimes even dominate
the usual decay modes. Decays to top squark pairs can similarly
be significant since $\tt_1$, the lighter of the two $t$-squarks,
can be much lighter than all other squarks.
These new decay channels lead to a diminution of the rate
into the standard decay modes so that the above-mentioned
``hole'' in parameter space becomes larger. Moreover, the region of
parameter space where more than one of the Higgs bosons is
visible above background (leading to unambiguous evidence
for a non-minimal Higgs sector) is substantially diminished.

One may also ask if the new supersymmetric decay modes of Higgs bosons
can lead to new avenues for detection. Decay modes of the neutral
Higgs bosons into $\tt_1\bar{\tt_1}$ and into $\tw_1\bar{\tw_1}$
will lead in general
to final states containing $0-2$ leptons plus jets plus $\eslt$; at the
proposed CERN Large Hadron Collider (LHC),
such signals are expected to be buried beneath standard model backgrounds
from processes such as top quark pair production. A more promising
signature\cite{BBDKT} may be found by searching for $H_p, H_h\to \tz_2\tz_2$,
where the neutralinos decay leptonically via $\tz_2\to \ell\bar\ell\tz_1$.
Such a process leads to a final state with as many as four isolated leptons
plus $\eslt$, which is expected to have small SM backgrounds. In addition,
kinematic information from the $4\ell +\eslt$ final state can yield
information on not only the Higgs boson masses, but also on the masses
of $\tz_2$ and $\tz_1$\cite{BBDKT}. Signals for invisible decays of
Higgs bosons via $H_\ell\to\tz_1\tz_1$ have also been considered\cite{JACK}.

In this paper, we seek to expand upon and improve the calculations
presented in Ref. \cite{BBDKT} regarding signals from supersymmetric
decays of the Higgs bosons. In particular,
\begin{itemize}
\item we incorporate radiative corrections to the SUSY Higgs
boson masses and couplings, from both top and bottom Yukawa interactions, and
effects from mixing between third generation squarks,
\item we include Higgs boson decays into squarks and sleptons---these
can be very important since decays to $\tt_1$ pairs may be allowed,
\item we have included Higgs production via $b\bar b$ fusion, which is the main
production mechanism for large values of $\tan\beta$,
\item we have included effects from non-degenerate squark and slepton masses,
which can lead to enhanced\cite{BT} leptonic decays of the
neutralino, resulting in a much larger
signal.
\end{itemize}

The rest of this paper is organized as follows. In Sec. 2, we describe in some
detail the improvements and extensions we have made to our earlier
calculation. In Sec. 3, we present cross-sections for the various
signals as a function of model parameters. In Sec. 4, we present
background calculations, and suggest a set of experimental cuts
useful for extracting the signal. We also estimate efficiencies
due to these cuts. In Sec. 5, we present our conclusions regarding
detectability
of the SUSY Higgs bosons via both the standard model decays as well
as via the supersymmetric $4\ell$ signal, along with a summary of our results.

\section{Calculational Details}

As in Ref. \cite{BBDKT}, we have computed the masses and mixing
angles in the Higgs boson sector using the 1-loop effective potential.
We have, however, improved on our previous calculation in several respects.
We now include the effects of both top and bottom family Yukawa
interactions (the latter interactions, which were neglected in
Ref. \cite{BBDKT}, can be important if $\tan\beta$ is large) as
well as effects from supersymmetric and SUSY-breaking trilinear scalar
interactions of the scalar top (scalar bottom) Higgs system. These
trilinear couplings lead to mixing between the L- and R- type sfermions,
and so further reduce the mass of the lighter mass eigenstate. This
can be especially important for $t$-squarks, whose mass is already
expected to be smaller than that of other squarks on account of the
large top Yukawa coupling, since the branching fraction for the
$\tt_1\bar{\tt_1}$ decay mode can be very large if $m_{\tt_1} < m_t$.
Finally, we have also included D-term contributions \cite{FN1} to the effective
potential, but generally speaking, their effects are small.

These modifications lead to improvements in the calculation of some of the
decay modes of the Higgs bosons.  First, radiative corrections at the same
level as those mentioned in the last paragraph are incorporated into the
calculation of the $H_h$-$H_{\ell}$-$H_{\ell}$ vertex.  Second, Higgs decays
into sfermions are now added.   As mentioned,
the sfermion masses including D-terms have been calculated with
intra-flavor mixing from non-zero values for $A_t$, $A_b$,
and $\mu$. These modifications to squark masses and mixings also
affect the loop decays of the Higgs bosons into two gluons which, in turn,
affects Higgs production by gluon fusion.

In the MSSM, the bottom quark Yukawa coupling is inversely
proportional to the parameter $\cos\beta$.
Hence, the $H\to b\bar b$ width (here, $H$ is a generic Higgs boson) as well
as the subprocess cross section $\hat{\sigma}(b\bar b\to H )$ are enhanced
in regions of large $\tan\beta$. In our calculations, we include Higgs
boson production cross sections via $gg$ and $b\bar b$ fusion. The formula for
$\hat{\sigma}(gg\to H)$ is well known\cite{CP}; for $b\bar b$ fusion, we
use the result
\begin{eqnarray}
\sigma (pp\to b\bar b\to HX) & = & {16\pi^2\over m_H^3}{1\over 4}{1\over 9}
{\Gamma (H\to b\bar{b})\over\lambda^{1\over 2}(1,m_b^2/m_H^2,m_b^2/m_H^2)}\tau
F_{DW}\times \\
& & \int_{\tau}^1 {dx\over x}
(D_{b/p}(x,Q^2) D_{\bar{b}/p}(x/\tau ,Q^2)+(b\leftrightarrow \bar{b} ))
\end{eqnarray}
where $\tau=m_H^2/s$, $D_{b/p}(x,Q^2)$ is the $b$ parton distribution function
in the proton, and
\begin{eqnarray}
F_{DW}=[16.2-4.28\ln (m_H)+0.31(\ln (m_H))^2]^{-1}
\end{eqnarray}
(with $m_H$ in GeV) is a fit to the results of Ref.\cite{DW} to incorporate
contributions of higher order graphs to the $b\bar b$ fusion mechanism. We use
the EHLQ Set 1\cite{EHLQ} $b$ and $g$ parton distributions, and take the
SM parameters to be $m_b = 5$ GeV, $\sin^2\theta_W = 0.23$,
$\alpha_s({M_Z}^2) = 0.118$ with $\Lambda_4(QCD) = 0.177$.

It has been pointed out in Ref. \cite{BT} that
leptonic decays of neutralinos can be enhanced by large factors if
squarks are significantly heavier than sleptons (as is the
case in the ``no-scale''\cite{LAHANAS} limit), and the $Z\tz_1\tz_2$
coupling is dynamically suppressed. This suppression is common in
supergravity models with radiative electroweak symmetry breaking since,
in this case, $|\mu |\simeq m_{\tg}$\cite{MSSM,SPECTRA}---as a result,
$\tz_1$ and $\tz_2$ are respectively mainly U(1) and SU(2) gauginos and
so have suppressed couplings to the Z boson. We calculate slepton masses
as usual by
using renormalization group equations to evolve sfermion masses from
a common GUT scale scalar mass to their weak scale values\cite{RGE}.
In our computation of the neutralino mass matrix we have, as usual,
assumed that the $\overline{MS}$ gaugino masses unify at some ultrahigh
energy scale. We then evolve these down to the electroweak scale. However, in
the following, we
present our results in terms of the physical pole gluino mass which we
relate to the corresponding $\overline{MS}$ mass
using the result of Ref.\cite{POLE}. Finally, we note that we have not
included QCD corrections to Higgs production and decay via loops\cite{ZERWAS}.

\section{The 4-lepton signal from neutralino decays of Higgs bosons }

Once the various Higgs boson production cross sections via $gg$ and $b\bar b$
fusion as well as the $H\to\tz_2\tz_2$ and
$\tz_2\to\ell\bar{\ell}\tz_1$ branching ratios are known, the  total rate for
$\sigma (pp\to H\to\tz_2\tz_2\to 4\ell +2\tz_1$) can be calculated. The cross
section is a function of the MSSM model parameters
($m_{\tg},m_{\tq},m_{\tt_L},m_{\tt_R}, m_{\tb_L}, \mu ,\tan\beta ,
m_{H_p}, A_t, A_b, m_t$); the slepton masses are related as given
in Ref. \cite{BT}. Of course, in
supergravity models with radiative electroweak symmetry breaking,
there is some correlation amongst these parameters: typically, $\mu$ scales
with $m_{\tg}$, and $m_{H_p}$ is strongly
correlated with $\mu$ and the universal GUT scale scalar mass. The parameter
choices used in this paper to illustrate our results are inspired,
but not ruled, by the supergravity mass relations.

We begin by showing in Fig. 1 a contour plot in the
$m_{H_p}$ vs. $\tan\beta$ plane
of $\sigma (pp\to H\to\tz_2\tz_2\to 4\ell+\eslt$ ($\ell =e$ or $\mu$)
in $fb$, at $\sqrt{s}=14$~TeV.
We take $m_{\tg}=-\mu=m_{\tq}=300$~GeV, while $A_t=A_b=0$, and
$m_t=165$~GeV. The scalar top masses are set at their default values:
$m_{\tt_L}^2 = m_{\tq}^2 - 50\ {\rm GeV}^2$, and
$m_{\tt_R}^2 = m_{\tq}^2 - 100\ {\rm GeV}^2$. The region in black is
excluded by the non-observation of supersymmetry signals in experiments
at LEP as parametrized in Ref.\cite{BBDKT} but taking into account the
recent limit of 63.5 GeV on the mass of the SM Higgs boson\cite{GOPAL}.
Fig. 1{\it a} shows results for $H=H_p$, while Fig. 1{\it b} shows
results for $H=H_h$. We note the following features.
\begin{itemize}
\item The $H_p\to 4\ell$ cross section exceeds $500$ $fb$ for small values of
$\tan\beta$ and $m_{H_p}\sim 300$~GeV, just below threshold for
$H_p\to t\bar t$ decay whereas the corresponding cross section from
$H_h$ does not exceed 100 $fb$. Over a wide range of parameters, away
from $\tan\beta=1$, the two
processes give comparable cross sections. The large difference between
the scalar and pseudoscalar contributions to the signal for small values
of $\tan\beta$ mainly comes from the difference in their SUSY branching
fractions\cite{BBDKT}.
\item The signal from the decays of $H_p$ and $H_h$
are separately larger than 5 $fb$ for 200 GeV $< m_{H_p} < 400$ GeV. Since
$H_h$ and $H_p$ are expected to be roughly degenerate over the
range of masses where the signal is significant, this corresponds
to a total of 250-25K $4\ell$ events at the LHC  before any selection
cuts, assuming a data sample with $50$ $fb^{-1}$. Interestingly,
the signal has a larger rate in the region below the dashed
line, where the lighter chargino is heavier than 90 GeV, which we take
to roughly represent the supersymmetry reach of LEP II (We note that
this region
is rather sensitive to the precise value of the chargino mass reach
that is assumed). The Higgs bosons $H_{\ell}$ and $H_p$ may themselves be
directly accessible at LEP II. The range of parameters where this is
possible is discussed in the last section and illustrated in Fig. 7
for one choice of parameters.
\item Although we have not shown this, we have checked that
for values of $\tan\beta\alt 4$, both $H_h$ and $H_p$
cross sections are dominated by gluon fusion, while $b\bar{b}$ fusion,
which was ignored in Ref.\cite{BBDKT}, dominates for
$\tan\beta \agt 10$.
\item The cross section for $4\ell +\eslt$ events has an observable
rate in part
of the ``hole region'', where $3<\tan\beta <10-20$ and
$100\alt m_{H_p}\alt 200-300$~GeV. This is the region where no SM decays
of MSSM Higgs are observable (except for possible observation of
$H\to b\bar b$). Of course, the range over which this signal might be
observable is sensitive to $\mu$ and $m_{\tg}$ since the cross section
drops sharply to zero when the kinematic limit for the
decays $H_{p,h}\to\tz_2\tz_2$ is approached.
\end{itemize}

The contours of the $4\ell$ cross section are shown in the $\mu-\tan\beta$
plane in Fig. 2 for ({\it a}) $H_p$ decays and ({\it b}) $H_h$ decays.
The pseudoscalar mass is fixed to be 250 GeV, which is within the hole region
for the light $\tw_1$ and $\tz_{1,2}$ case illustrated here. Other
parameters are as in Fig. 1. We see that the signal cross section exceeds
10 $fb$ for a wide region of parameter space. Again, the black region
is excluded by LEP constraints and the dashed line is the contour
$m_{\tw_1}=90$ GeV (the region above the dashed contour has
$m_{\tw_1}<90$ GeV). In the upper corners of Fig. 2,
$m_{\tilde{\tau}}< 45$ GeV, because of $\tau$
Yukawa interactions which have substantial effects for large
values of $\tan\beta$\cite{DN}.
While the signal is small
for small values of $|\mu |$, it is
instructive to see that the $4\ell$ rate appears
to be observable for $|\mu |\simeq m_{\tg}$ as expected in
supergravity models with radiative EW symmetry breaking.
Finally, we observe that the pseudoscalar Higgs
boson tends to give a larger signal than the heavy scalar, and further,
that the signal is significantly larger if $\mu$ happens to be negative in
our convention.

In Fig. 3, we study the dependence of the signal on the squark mass for a
fixed value of $m_{\tg}$. Here, $m_{H_p}$ is fixed at 250 GeV; other
parameters are as in Fig. 1.
We see that the $4\ell$ cross section rapidly
drops off as $m_{\tq}$ increases from $m_{\tg}$ to larger values. For
values of $\tan\beta \agt 5$, the cross section shows a
slow increase for rather large squark masses. As expected,
the squark mass dependence
of the cross section is mostly determined by that of the leptonic branching
fraction of the $\tz_2$. For $m_{\tq}\simeq m_{\tg}$, the sleptons
are considerably lighter than the squarks resulting\cite{BT} in an
enhancement of $\tz_2\to\ell\bar{\ell}\tz_1$ decays, and hence, in the $4\ell$
cross section. The increase in the cross section for large values
of $m_{\tq}$ comes from interference between various amplitudes.
We see that the signal has an observable rate even if squarks are
significantly heavier than gluinos. Amusingly, the cross section is
largest below the dashed line where the chargino is likely to be undetectable
at LEP II.

Before closing this discussion we mention that we have checked that the $4\ell$
cross section is relatively insensitive to the $A$-parameters or to the
precise value of $m_t$ except when the threshold for a new decay is crossed.
This is especially important when considering the variation of the signal
with $A_t$, since $H\to\tt_1\bar{\tt_1}$ decays become kinematically
accessible when $|A_t|$ becomes very large. Finally, we have also studied
the variation of the signal with $m_{\tg}$. For $m_{H_p} = 250$ GeV, and
$-\mu = m_{\tq} = m_{\tg}$, the signal exceeds 5 $fb$ for gluinos as heavy
as 350-400 GeV. This is reasonable since then, $m_{\tz_2} \simeq = m_{\tg}/3$
is approaching the kinematic boundary of $H_p\to \tz_2\tz_2$ decays.

\section{Backgrounds, Cuts and Efficiencies}

We have seen that the neutralino
decays of the pseudoscalar and the heavy neutral scalar Higgs bosons of
the MSSM can result in several hundred to several thousand events with
four isolated leptons together in
an LHC data sample of 50 $fb^{-1}$. These events should be very distinctive
since, except for QCD radiation, they would be free of hadronic activity.
Within the SM, isolated leptons mainly arise from the the production and
decays of the $W$ and $Z$ bosons and of the top quark. The main background
from $ZZ$ pair production can be efficiently removed by vetoing events
where like-flavour, opposite sign leptons reconstruct the $Z$ mass within
$\pm 10$ GeV. We have verified this by generating 10K $ZZ$ events
using ISAJET (and forcing the leptonic decay of the Z bosons). No
events are found after the mass cut, resulting in an upper bound
on the $4\ell$ cross section of about 0.003 $fb$.
This would, of course, also remove the signal if the decay
$\tz_2\to\tz_1 Z$ is kinematically accessible; fortunately for the range
of parameters where we find the signal to be substantial, this decay
is kinematically inaccessible, so that this requirement results in
very little loss of the signal. The signal can also be mimicked
by $\bar{t}tZ$ or $4t$ production. These backgrounds can easily be
removed by vetoing events with a central jet in addition to the
$Z$-mass veto already mentioned. The main SM physics
background thus comes from electroweak multi-$W$ production. At the LHC,
the trilepton cross section from $3W$ production has been shown\cite{BH} to be
about 2 $fb$, so that the background to the signal from $4W$ production should
be negligible. Because the cross section for $t\bar{t}$ production at
the LHC is very large, one may also worry that these may provide a significant
background when the leptons from the daughter bottom quarks are accidently
isolated; a simulation of 300K $t\bar{t}$ events with forced top quark decays
yielded no background,
giving a limit $\sigma (t\bar t )<4\ fb$.
We have been unable to identify any significant SM sources of
physics backgrounds to the SUSY Higgs boson signal.

Within the MSSM framework, however, the continuum production of $\tz_2$
pairs\cite{FN3}
can also result in the same signal. These are produced by $q\bar{q}$
annihilation via $s$-channel $Z$ exchange or $t$- and $u$-channel squark
exchange. While the detection of these continuum neutralino pairs would
in itself
be very exciting, it is interesting to ask whether $\tz_2\tz_2$ production
via $H_h$ and $H_p$ decays is distinguishable from the continuum production
of $\tz_2$ which, in effect, is the background to our Higgs signal. Toward
this end, we have shown in Fig. 4 the cross section for $4\ell$ production via
continuum $\tz_2\tz_2$ production as a function of $\tan\beta$.
To compare with $\tz_2$ production via Higgs boson decays discussed
in the last section, we have taken $m_{\tg}=\pm\mu=300$ GeV and
illustrated our results for
$m_{\tq}=m_{\tg}$ (solid line) and $m_{\tq}=2m_{\tg}$ (dashed line).
The pseudoscalar Higgs boson mass is fixed to be 250 GeV and other
parameters are fixed as in Fig. 1-3.
We see that while the cross section is sensitive to the squark mass
it is relatively insensitive to $\tan\beta$ over the region
where the cross section is significant. Furthermore, for $m_{\tg} = m_{\tq}$,
the cross section is 10-30 $fb$ which is generally
comparable to, or smaller than,
the cross sections in Fig. 2 for a wide range of parameters.

We use ISAJET 7.07\cite{ISAJET} to
simulate the $4\ell$ Higgs boson signal. Since explicit SUSY Higgs
production has not yet been incorporated into this code, we simulate the
production of $H_h$ or $H_p$ by decaying the SM Higgs scalar into a
$\tz_2$ pair and forcing the SUSY decay mode; the total cross section is then
normalized to the results of Fig. 1-3.

We use the toy calorimeter simulation package ISAPLT to model detector effects.
We simulate calorimetry with cell size
$\Delta\eta\times\Delta\phi =0.1\times 0.1$, which extends between
$-5<\eta <5$ in pseudorapidity. We take electromagnetic
energy resolution to be $10\% /\sqrt{E_T}\ \oplus\ 0.01$, while hadronic
resolution is $50\% /\sqrt{E_T}\ \oplus\ 0.03$ for $|\eta | <3$, and
$100\% /\sqrt{E_T}\ \oplus \ 0.07$ for $3<|\eta |<5$, where $\oplus$
denotes addition in quadrature.
Jets are coalesced
within cones of $R=\sqrt{\Delta\eta^2 +\Delta\phi^2} =0.7$ using
the ISAJET routine GETJET. Hadronic clusters with $E_T>50$ GeV
are labelled as jets.
Muons and electrons are classified as isolated if they have $p_T>10$ GeV,
$|\eta (\ell )|<2.5$,
and the visible activity within a cone of $R =0.3$ about the lepton
direction is less than $E_T({\rm cone})=2$ GeV.

We then impose the following cuts designed to select signal events, while
vetoing SM backgrounds from $ZZ$ and $t\bar t$
production:
\begin{itemize}
\item require {\it two} isolated leptons with $p_T(\ell )>20$ GeV to trigger
the event,
\item require {\it two} more isolated leptons with $p_T(\ell )>10$ GeV,
\item require all opposite sign but same flavor dilepton pairs to
have invariant mass $m(\ell\bar\ell )<80$ GeV or $m(\ell\bar\ell )>100$ GeV,
\item require number of jets $n({\rm jets})=0$.
\end{itemize}
A cut on $\eslt$ could be considered instead of the above dilepton mass cut.
However, the $\eslt$ spectrum from the signal is not so hard, while
forward jet production and energy mis-measurement in $ZZ$ events can lead
to substantial $\eslt$, so that the dilepton mass cut wins over an $\eslt$
cut in rejecting background while preserving signal.

We should stress that there are no significant SM backgrounds even before
the last cut. Backgrounds involving the $Z$ are efficiently removed by
the $m(\ell\bar\ell )$ cut. The main physics background would then
be expected to come from $4t$ production which gives a $4\ell$ cross
section around 0.05$fb$ at the LHC, even before acceptance cuts\cite{BSP}.
The last cut which removes about 40\% (70\%) of the signal for a Higgs
mass around 200 GeV (400 GeV) has been imposed to separate neutralino
production from the cascade decays of gluinos and squarks\cite{BTW} which
can also produce multilepton events at observable rates.

In order to give the reader an idea of the impact of the cuts on the $4\ell$
signal, we have shown these
cross sections for illustrative choices of $m_{H_p}$ and $\tan\beta$ in Table
I.
We have added the contributions from $H_h$ and $H_p$ decays since these
are expected to lead to kinematically similar events.
We have also shown the continuum background for the same choices of parameters.
The following points are worth noting.
\begin{itemize}
\item For the choice of parameters in the Table, the signal exceeds the
background for $m_{H_p}$ up to somewhat beyond 300 GeV, where $t\bar{t}$ decays
become accessible. Also, the cross section
corresponds to an event rate 100-1000 events after cuts in a 50 $fb^{-1}$ data
sample, compared to a continuum background of 50-100 events.
\item The signal efficiency varies between 5 and 10 percent depending on
the model parameters; Fig. 1-3 can thus be used to estimate the signal
after the cuts.
\end{itemize}

It should, of course, be kept in mind that the SUSY parameters are not
known, and that the total background (and signal) rate
could be considerably different
(even for similar values of $m_{\tz_2}$)
from our estimate in the Table.
An excess of the $4\ell$ events relative to the background in Table I
would, therefore, not enable us to cleanly infer a Higgs boson
signal. Instead, we consider the possibility of separating the signal
from the continuum background by studying the invariant mass distribution of
the four leptons: for the signal, we must have $m$($4\ell$)$<m_H-2m_{\tz_1}$,
while the background should exhibit a rather broad distribution. Since these
distributions are determined by the Higgs boson and neutralino masses, we
expect them to be relatively insensitive to variations in model
parameters which result in similar values of $m_H$, $m_{\tz_1}$ and
$m_{\tz_2}$.

Toward this end, we have shown in Fig. 5
and Fig. 6
these distributions for the signal plus background (solid) and the $\tz_2\tz_2$
continuum ``background'' (dashed) for the six cases in Table I. For the
smaller values of $m_{H_p}$, the solid histograms are dominated by the signal
and differ considerably from the dashed background histograms (note that
these are shown using a log scale). As anticipated above, the solid and dashed
lines indeed coincide for $m(4\ell )>m_{H_p}-2m_{\tz_2}$. In order to decide
whether the solid and dashed histograms are indeed distinguishable, we have
computed the total $\chi^2$ for the difference
between these two histograms,
{\it normalized to the same number of events}, with the event number given
by the signal plus background cross section times an integrated luminosity
of 50 $pb^{-1}$. In this computation, we have
used twelve 20 GeV bins between $m(4\ell)=60$ GeV and $m(4\ell)=300$ GeV.
In our simulation, we have about 1100, 900 and 600 events for
$m_{H_p}=200,\ 300$ and 400 GeV (after cuts) for the
$\tan\beta=2$ case, and about 550 events for each value of $m_{H_p}$ for
$\tan\beta =10$.
The resulting total $\chi^2$ is shown in the last column of Table I. It thus
appears that for $m_{H_p}\alt 300$ GeV, the distributions are
sufficiently different that the solid line is unlikely to be a chance
fluctuation of the continuum background (for $\chi^2 >26.2$, this probability
is
smaller than 1\%). Some remarks are, however, in order:

\begin{itemize}
\item Despite the fact that we have normalized the solid and histograms
to have the same number of events, our conclusion clearly depends on the
relative number of Higgs initiated and continuum $\tz_2\tz_2$ events. Here,
we have used the rate as given by the MSSM for the values of parameters
motivated by supergravity models.
\item For the first two cases in Table I, the number of events in our
simulation is comparable to the expected number in a data sample of
50 $fb^{-1}$. We thus expect our computation of the total $\chi^2$ to be a
reasonable
reflection of the experimental situation. For the $\tan\beta$=10 case, with
$m_{H_p} = 200$ GeV (300 GeV),
the signal cross section is much smaller so that about 200 $fb^{-1}$
(400 $fb^{-1}$) of integrated luminosity need to be collected in order
that the fluctuations in our simulation are of comparable magnitude to
those in the data. We thus conclude that while 50 $fb^{-1}$ of data may
suffice to enable one to
distinguish the Higgs signal from continuum $\tz_2\tz_2$ production
for smaller
values of $\tan\beta$,
integrated luminosities of
200-400 $fb^{-1}$
may be necessary if $\tan\beta$ is large.
\item We should bear in mind that we have used only the shape of the $4\ell$
mass distribution to try to untangle the Higgs signal from continuum
$\tz_2\tz_2$ production without any regard for rate or other event shape
variables. It would be interesting to explore whether other distributions
serve as better discriminators of the Higgs from the continuum background.
\end{itemize}

\section{Concluding Remarks}

If low energy supersymmetry is to provide a rationale for the stability
of the electroweak scale, sparticles must all be lighter than about 1 TeV.
In models where gaugino masses are unified at an ultra-high unification
scale, the lighter chargino and the two lightest neutralinos are frequently
lighter than 100-150 GeV, and are thus expected to be kinematically accessible
in the decays of the heavier Higgs bosons of the model. In a previous
paper\cite{BBDKT}, we had shown within the MSSM framework that these
SUSY modes dominated the decays of the heavier neutral Higgs boson,
and particularly, the pseudoscalar Higgs boson over a wide range of SUSY
parameters. This has two important consequences. The down side is that
it reduces the cross sections for the $\gamma\gamma$ and $ZZ$ or $ZZ^*$
modes which form the usual signal for Higgs bosons at hadron colliders. These
SUSY decays, however, open up new possibilities for Higgs boson detection, the
most promising of which is the $4\ell$ signal from the $\tz_2\tz_2$ decays of
$H_p$ or $H_h$. Here, we have improved on our previous computation of this
signal, and also explored how it varies with model parameters. In this
connection, we have used supergravity models as a guide in order to
restrict the parameter space, and make this exploration tractable.

For the convenience of the reader, and also because the projected energy
of the LHC has been reduced to 14 TeV, we first briefly review the
detectability
of the SM decay modes of the Higgs boson at the LHC. For the $\gamma\gamma$
signal, we require the center of mass scattering angle satisfy
$\cos\theta^* < 0.8$\cite{fn2}. As before\cite{BBKT}, the background is assumed
to come from continuum pair production via
$q\bar{q},gg\to\gamma\gamma$ where the photons
have $m_{\gamma\gamma}=m_H\pm 1\%$. For the $ZZ$ and $ZZ^*$ signal, we have
required that the four leptons reconstruct to the Higgs boson mass within
a mass-dependent resolution given in Ref.\cite{BBDKT}. We have considered
the $ZZ^*\to 4\ell$ signal only for $m_H>130$ GeV, in which case backgrounds
from continuum $ZZ^*$ and $Z\gamma^*$ production
have been shown to be negligible\cite{SDC}.

The regions of the $m_{H_p}-\tan\beta$ plane where these various signals are
observable using the 99\% CL criterion described in Ref.\cite{BBKT} is
illustrated in Fig. 7 for ({\it a}) $m_{\tg}=m_{\tq}=-\mu=1$ TeV, and
({\it b}) $m_{\tg}=m_{\tq}=-\mu=-300$ GeV.
In this figure, we have assumed $A_t=A_b=0$ and taken the
integrated luminosity to be 50 $fb^{-1}$. The legends in this figure appear on
the side of the boundary where the signal is observable. The black region
denotes the range of parameters excluded by experiments at LEP\cite{GOPAL}
while the regions below the lines marked LEP190 and LEP175 can be probed in
experiments at LEP with optimistic ($\sigma(H_{\ell}H_p)$ or
$\sigma(H_{\ell}Z)>0.05$ $pb$, $\sqrt{s}=190$ GeV) and pessimistic
($\sigma(H_{\ell}H_p)$ or $\sigma(H_{\ell}Z)>0.2$ $pb$, $\sqrt{s}=175$ GeV)
scenarios for performance of the LEP collider in
its next phase. We note that:
\begin{itemize}
\item We have used an integrated luminosity of 50 $fb^{-1}$ because
at the reduced energy, we found that the $H_{\ell}\to\gamma\gamma$ signal was
essentially unobservable over the whole plane in Fig. 7b with a data
sample of ``just'' 30 $fb^{-1}$. This is a reflection of the well-known
fact that
the position of this contour is extremely sensitive to
the assumptions about the detector.
\item The shaded region is where none of the neutral Higgs bosons of the
MSSM are detectable at either the LHC or at LEP II, at least via the signals
discussed above. As noted in the Introduction, it may be possible to detect
Higgs boson signals even for parameters inside this ``hole'' if Higgs decays
to $\tau$-leptons\cite{KZ} or bottom quark pairs\cite{DGV} are identifiable.
\item It is amusing to see that the shaded region actually shrinks in
Fig. 7b. This is somewhat misleading because this shrinkage is due to
the upward movement of the LEP190 curve. Conservative assumptions
about the performance of LEP considerably increase the region where
there is no signal either at LEP II, or at the LHC. We have also
checked that increasing the value of the $A$-parameter lowers the
LEP  190 curve. We have traced this to an increase in $m_{H_{\ell}}$, and
the corresponding suppression of the cross section for $ZH_{\ell}$ production:
for $A_t=400$ GeV or $A_t=-700$ GeV, the LEP 190 curve roughly follows the
LEP175 curve in the figure. {\it Thus, the ``hole" in Fig. 7{\it b} may well be
considerably bigger than indicated by the shaded region even with optimistic
assumptions about the performance of LEP II.}
\item We note here that the main reason for the change in the
$H_{\ell}\to \gamma\gamma$
boundary when the parameters are altered from 1 TeV to 300 GeV is {\it not}
the opening of the SUSY decays of $H_{\ell}$. The shift occurs primarily
because the Higgs mass is altered (due to the incorporation of the
improved radiative corrections) resulting in a different size of the
background.
\item Finally, we note that as anticipated, the region of parameters
where the signal from two different Higgs bosons is simultaneously
detectable is greatly reduced for the case in Fig. 7b
\end{itemize}

Turning to the neutralino signal for Higgs bosons, the cross sections for the
$4\ell$ signal from $\tz_2\tz_2$
decays of the Higgs bosons are summarized in Figs. 1-3. We see that these
cross sections can be as large as 500 $fb$ for ranges of model parameters
allowed by all known experimental data. It is also worth noting that, before
experimental cuts,
the signal exceeds 10 $fb$ over a large region of parameter space where
there may be no visible SUSY signal even at LEP II. We have argued that
there are no significant SM backgrounds to this signal. Thus while the
detection of four lepton events at such rates will be a signal for new
physics, its identification as a Higgs boson signal requires that it be
separable from continuum $\tz_2\tz_2$ production for which the cross section
is shown in Fig. 4, for $\mu=\pm m_{\tg}$ as expected
in supergravity models. We see that over a large range
of parameters, this background is significantly smaller than the signal
in Fig. 1-3. The efficiency with which this signal may be detected
is shown in Table I, for cuts typical of LHC detectors.
We see that this is typically 5-10\%, so that the cross sections in Fig. 1-3
correspond to 25-1000 events
in an experimental data sample of 50 $fb^{-1}$.

In order to assess whether the Higgs signal could be distinguished above
the neutralino continuum, we examined the mass distribution of the four
leptons produced via the decay of the Higgs boson. For illustrative values
of model parameters, we showed that the shape of this distribution
may serve to discriminate the Higgs signal from continuum neutralino
production provided $m_{H_p}\alt 2m_t$. We have argued
that for low values of $\tan\beta$ an integrated luminosity of 50 $fb^{-1}$
is sufficient for this discrimination, but an integrated luminosity of
200-400 $fb^{-1}$ is required if $\tan\beta$ = 10. We have, respectively,
denoted these cases by crosses and open circles in Fig. 7b. As we can see
SUSY decays of Higgs bosons are indeed detectable well into the ``hole''
region if LEP II is operated at about 175 GeV. Although this is
not obvious from the figure, this may also be the case with optimistic
assumptions about the performance of LEP II, since, as we mentioned
the LEP II observability curve essentially follows the LEP 175 curve
if $|A_t|$ is large, while the Higgs signal, is relatively
insensitive to variations in $A_t$.

In summary, we have shown that the processes
$H_{h,p}\to\tz_2\tz_2\to 4\ell +\eslt$
lead to an observable rate for gold-plated four lepton events at the LHC for a
significant range of SUSY model parameters. The parameter space region
where the neutralino decays of SUSY Higgs bosons ought to be
observable can be summarized as
\begin{itemize}
\item $2m_t>m_{H_p}>2m_{\tz_2} \simeq 2m_{\tg}/3$,
\item $m_{\tq}\sim m_{\tg}$ so that $m_{\tl}<<m_{\tq}$
and $|\mu|\sim m_{\tg}\alt 500$ GeV
\item $m_{H_p}\sim 200-350$ GeV,
\end{itemize}
We note that if SUSY parameters are in this region, there will also be
a plethora of other signals via which SUSY will be detectable at the
LHC\cite{BTW}).

We have been unable to identify any
significant SM background to this distinct SUSY Higgs boson
signal. We have shown that
this signal may be distinguished from continuum $\tz_2\tz_2$ production
provided a sufficiently large integrated luminosity is available. Finally,
this process may provide the only way to identify any Higgs boson of the
MSSM if the model parameters are in the ``hole'' region (unless identification
of their $\tau$
and bottom quark decays turns out to be feasible), and further, that it
may well provide the only identifiable signal for the notoriously hard
to detect pseudoscalar Higgs boson.

%
\acknowledgments

This research was supported in part by the U.~S. Department of Energy
under contract number DE-FG05-87ER40319 and DE-AM03-76SF00235.
In addition, the work of HB was supported by the TNRLC SSC Fellowship program.
%
%
%
%

%
\newpage
%
%

\begin{table}
\caption[]{Cross sections in fb at LHC for $4\ell +\eslt$ events from
supersymmetric processes. We take $m_{\tg}=300$~GeV and $\mu=-m_{\tg}$,
while $m_{\tq}\sim m_{\tg}$ and $A_t =0$.}
\bigskip
\begin{tabular}{lrrrrrrrr}
process & $m_{H_p}$ & $\tan\beta$ & $m_{\tz_2}$ & $m_{\tz_1}$ &
$\sigma (4\ell )$ & $\sigma (cut)$ & ${\rm effic.}$ & $\chi^2$ \\
\tableline
$H_h,H_p\to 4\ell$ & 200 & 2 & 97.3 & 45.5 & 230 & 26 & 11\% & 18036 \\
$H_h,H_p\to 4\ell$ & 300 & 2 & " & " & 260 & 23 & 9\% & 1024 \\
$H_h,H_p\to 4\ell$ & 400 & 2 & " & " & 37 & 2.4 & 6.5\% & 23 \\
$\tz_2\tz_2\to 4\ell$ & -- & 2 & " & " & 27 & 2.8 & 10\% & -- \\
$H_h,H_p\to 4\ell$ & 200 & 10 & 83.4 & 42.8 & 44 & 2.4 & 5.5\% & 528 \\
$H_h,H_p\to 4\ell$ & 300 & 10 & " & " & 26 & 1.4 & 5.4\% & 32 \\
$H_h,H_p\to 4\ell$ & 400 & 10 & " & " & 10 & 0.5 & 5\% & 4.5 \\
$\tz_2\tz_2\to 4\ell$ & -- & 10 & " & " & 10 & 0.8 & 8\% & -- \\
\end{tabular}
\end{table}



%
\begin{figure}
\caption[]{Contour plot of cross section in $fb$ for
{\it a}) $pp\to H_p\to\tz_2\tz_2\to 4\ell +\eslt$ and
{\it b}) $pp\to H_h\to\tz_2\tz_2\to 4\ell +\eslt$ events, in the
$m_{H_p}$ vs. $\tan\beta$ plane, at $\sqrt{s}=14$~TeV. We take
$m_{\tq}=m_{\tg}=-\mu =300$~GeV, $m_t=165$~GeV, and $A_t=A_b=0$.
We also take $m_{\tt_L}^2=m_{\tq}^2-50\ {\rm GeV}^2$ and
$m_{\tt_R}^2=m_{\tq}^2-100\ {\rm GeV}^2$.
The region above the dashed lines corresponds to $m_{\tw_1}<90$~GeV, the
approximate reach of LEP II. The reach of LEP II for detection of Higgs
boson signals is shown in Fig. 7b}
\end{figure}
%
\begin{figure}
\caption[]{Contour plot of cross section in $fb$ for
{\it a}) $pp\to H_p\to\tz_2\tz_2\to 4\ell +\eslt$ and
{\it b}) $pp\to H_h\to\tz_2\tz_2\to 4\ell +\eslt$ events, in the
$\mu$ vs. $\tan\beta$ plane, at $\sqrt{s}=14$~TeV.
Parameters are as in Fig.~1, except $m_{H_p}=250$~GeV.
The region above the dashed contour corresponds to $m_{\tw_1}<90$ GeV.}
\end{figure}
%
\begin{figure}
\caption[]{Contour plot of cross section in $fb$ for
{\it a}) $pp\to H_p\to\tz_2\tz_2\to 4\ell +\eslt$ and
{\it b}) $pp\to H_h\to\tz_2\tz_2\to 4\ell +\eslt$ events, in the
$m_{\tq}$ vs. $\tan\beta$ plane, at $\sqrt{s}=14$~TeV.
Parameters are as in Fig.~1, except $m_{H_p}=250$~GeV.
The region above the dashed contour corresponds to $m_{\tw_1}<90$ GeV.}
\end{figure}
%
\begin{figure}
\caption[]{Cross section in $pb$ for continuum
$pp\to\tz_2\tz_2\to 4\ell +\eslt$ production versus $\tan\beta$ for
$m_{\tg}=\pm \mu =300$~GeV, at $\sqrt{s}=14$~TeV.}
\end{figure}
%
\begin{figure}
\caption[]{Distribution in $m(4\ell )$ after cuts
from signal plus background (solid),
and background (dashes) for $pp\to H_p,H_h\to\tz_2\tz_2\to 4\ell +\eslt$
production for $\tan\beta=2$ for {\it a}) $m_{H_p}=200$ GeV,
{\it b}) $m_{H_p}=300$ GeV, and {\it c}) $m_{H_p}=400$ GeV. Other
parameters are as in Fig. 1.}
\end{figure}
%
\begin{figure}
\caption[]{Distribution in $m(4\ell )$ after cuts
from signal plus background (solid),
and background (dashes) after cuts
for $pp\to H_p,H_h\to\tz_2\tz_2\to 4\ell +\eslt$
production for $\tan\beta=10$ for {\it a}) $m_{H_p}=200$ GeV,
{\it b}) $m_{H_p}=300$ GeV, and {\it c}) $m_{H_p}=400$ GeV. Other
parameters are as in Fig. 1.}
\end{figure}
%
\begin{figure}
\caption[]{Plot of discovery regions in the
$m_{H_p}$ vs. $\tan\beta$ plane, at $\sqrt{s}=14$~TeV, assuming an
integrated luminosity of $50\ fb ^{-1}$. In {\it a}), we take
$m_{\tq}= m_{\tg}=-\mu =1000$~GeV, while in {\it b}), we
take $m_{\tq}= m_{\tg}=-\mu =300$~GeV, so that SUSY decay of Higgs
bosons are allowed. Other parameters are as in Fig. 1. The Higgs boson
reach of LEP II is sensitive to the value of the $A$-parameter. For large
values of $|A|$, the LEP 190 curve roughly follows the LEP 175 curve as
discussed in the text.}
\end{figure}
%


\begin{references}
\bibitem{FN1} The D-term corrections are proportional to $f_t^2g^2$ or
$f_b^2g^2$, where the $f_i$ denote the Yukawa couplings and $g$, the
gauge couplings. Hence, their inclusion without the inclusion of pure
gauge loops with contributions proportional to $g^4$ is formally consistent.
\bibitem{FN3} For $m_{\tg}=-\mu$, $\tz_{3,4}$ decay via two body modes
so that their continuum production will not be confused with the signal.
The same is frequently the case for other values of $\mu$.
\bibitem{fn2} In our previous calculations, the $\gamma\gamma$ background
had been under estimated since the photons were inadvertently required
to satisfy $\cos\theta^* < 0.5$ instead of 0.8.
\end{references}
\end{document}